\documentclass
[aps,floatfix,twocolumn,byrevtex,superscriptaddress]{revtex4-1}
\usepackage{amsmath}
\usepackage{amssymb}
\usepackage{amstext}
\usepackage{amsopn}
\usepackage{amsfonts}
\usepackage{amsxtra}
\usepackage[english]{babel}
\usepackage{graphicx}

\usepackage{float}
\usepackage{bm}
\usepackage{multirow}
\usepackage{dcolumn}
\usepackage{color}
\usepackage{hyperref}

\newcommand{\icm}{\ensuremath{~\textrm{cm}^{-1}}}

\begin{document}

\title{Band-selective clean- and dirty-limit superconductivity with nodeless gaps in the bilayer iron-based superconductor CsCa$_2$Fe$_4$As$_4$F$_2$}

\author{B. Xu}
\email[]{bing.xu@unifr.ch}
\affiliation{University of Fribourg, Department of Physics and Fribourg Center for Nanomaterials, Chemin du Mus\'{e}e 3, CH-1700 Fribourg, Switzerland}

\author{Z. C. Wang}
\affiliation{Department of Physics and State Key Lab of Silicon Materials, Zhejiang University, Hangzhou 310027, China}

\author{E. Sheveleva}
\author{F. Lyzwa}
\author{P. Marsik}
\affiliation{University of Fribourg, Department of Physics and Fribourg Center for Nanomaterials, Chemin du Mus\'{e}e 3, CH-1700 Fribourg, Switzerland}

\author{G. H. Cao}
\affiliation{Department of Physics and State Key Lab of Silicon Materials, Zhejiang University, Hangzhou 310027, China}

\author{C. Bernhard}
\email[]{christian.bernhard@unifr.ch}
\affiliation{University of Fribourg, Department of Physics and Fribourg Center for Nanomaterials, Chemin du Mus\'{e}e 3, CH-1700 Fribourg, Switzerland}

\date{\today}
%
%

\begin{abstract}
The optical properties of the new iron-based superconductor CsCa$_2$Fe$_4$As$_4$F$_2$ with $T_c \sim 29$~K have been determined. In the normal state a good description of the low-frequency response is obtained with a superposition of two Drude components of which one has a very low scattering rate (narrow Drude-peak) and the other a rather large one (broad Drude-peak). Well below $T_c \sim 29$~K, a pronounced gap feature is observed which involves a complete suppression of the optical conductivity below $\sim$ 110~cm$^{-1}$ and thus is characteristic of a nodeless superconducting state. The optical response of the broad Drude-component can be described with a dirty-limit Mattis-Bardeen-type response with a single isotropic gap of $2\Delta \simeq 14$~meV. To the contrary, the response of the narrow Drude-component is in the ultra-clean-limit and its entire spectral weight is transferred to the zero-frequency $\delta(\omega)$ function that accounts for the loss-free response of the condensate. These observations provide clear evidence for a band-selective coexistence of clean- and dirty-limit superconductivity with nodeless gaps in CsCa$_2$Fe$_4$As$_4$F$_2$.
\end{abstract}



\maketitle

%
%
The discovery of high-temperature superconductivity in the iron-based superconductors (FeSCs) has received great attention over the past decade~\cite{Kamihara2008,Chen2008,Chen2008PRL,Paglione2010,Stewart2011}. Meanwhile, a large number of different families of FeSCs have been discovered which all have FeAs or FeSe layers as their essential structural element. According to the stacking of these FeAs or FeSe layers and the additional layers that separate them, they can be classified into 1111-type (e.g. LaFeAsO)~\cite{Kamihara2008,Chen2008,Chen2008PRL}, 111-type (e.g. LiFeAs)~\cite{Wang2008}, 11-type (e.g. FeSe)~\cite{Hsu2008}, 122-type (e.g. BaFe$_2$As$_2$)~\cite{Rotter2008a,Rotter2008}, etc. Probably the most intensively studied is the 122-type family for which large and high-quality single crystals can be readily grown. The 122 parent compounds are antiferromagnetic metals for which superconductivity is obtained by applying chemical doping, which leads to extra holes~\cite{Rotter2008}, electrons~\cite{Athena2008} or chemical pressure (for isovalent substitution)~\cite{Kasahara2010}, or by applying external pressure~\cite{Alireza2008}. Recently, the closely related CaKFe$_4$As$_4$ (1144-type)~\cite{Akira2016} family with $T_c \sim 35$~K, has been discovered. Unlike K-doped Ca-122, for which the K and Ca ions are randomly distributed among the layers that separate the FeAs planes, the 1144-type structure is composed of distinct K and Ca layers that alternatingly separate the FeAs planes. Accordingly, the 1144-structure has two well-defined As sites that are neighboring either a K or a Ca layer. This strongly reduces the disorder and it also introduces an electric field gradient across the FeAs planes that can, e.g., affect the orbital state of the charge carriers, enhance spin-orbit coupling effects and even affect the antiferromagnetic order~\cite{Meier2018}. Very recently, the so-called 12442-type (KCa$_2$Fe$_4$As$_4$F$_2$) family with $T_c \sim 30$~K in its stoichiometric form was successfully synthesized for which the Ca layer of CaKFe$_4$As$_4$ is replaced with a Ca$_2$F$_2$ layer~\cite{Wang2016,Wang2017}. This should further enhance the electric field gradient across the FeAs layers and strongly increase the anisotropy of the electronic and superconducting (SC) properties in the directions parallel (ab-plane) and perpendicular (c-axis) to the FeAs layers. The intrinsic hole doping of this 12442-type compound corresponds to the one of a 50 percent K-doped Ba-122 sample placing it in the moderately overdoped regime of the phase diagram where the $T_c$ values remain sufficiently high and competing magnetic and/or structural orders are absent.

The new 12442-type family thus is well-suited to study which effects the reduced disorder, a strong anisotropy, and an enhanced spin-orbit coupling due to broken local inversion symmetry of the FeAs planes have on the unconventional superconducting properties of the FeSCs. Recent $\mu$SR measurements on polycrystalline samples of the 12442 compounds KCa$_2$Fe$_4$As$_4$F$_2$~\cite{Smidman2018} and CsCa$_2$Fe$_4$As$_4$F$_2$~\cite{Kirschner2018} have indeed provided evidence for the presence of line nodes in the SC gaps. This result is very surprising and in clear contrast with the experiments on CaKFe$_4$As$_4$ single crystals which revealed multiple Fermi surfaces with nodeless SC gaps~\cite{Mou2016,Cui2017,Biswas2017,Cho2017,Yang2017}, in good agreement with previous results on optimally doped Ba$_{1-x}$K$_x$Fe$_2$As$_2$~\cite{Ding2008}. These contradictory results thus call for further investigations especially of the SC gap structure of high-quality 12442 single crystals.

In this work, we present such a study based on optical spectroscopy measurements of high-quality CsCa$_2$Fe$_4$As$_4$F$_2$ single crystals. The normal state optical properties of this multiband material are analyzed with a two-Drude model which reveals two types of free carrier that can be distinguished according to their very small and rather large scattering rates, respectively. In the superconducting state, our optical spectra provide clear evidence for the absence of any nodal gaps. Instead, the optical conductivity is well described with two isotropic gaps that are in the extremely-clean and the dirty limit, respectively.

%
%
High-quality single crystals of CsCa$_2$Fe$_4$As$_4$F$_2$ were grown using the self-flux method with a CsAs flux~\cite{Wang2018}. The \emph{ab}-plane reflectivity $R(\omega)$ was measured at a near-normal angle of incidence with a Bruker VERTEX 70v FTIR spectrometer with an \emph{in situ} gold overfilling technique~\cite{Homes1993}. Data from 40 to 12\,000\icm\ were collected at different temperatures from 300 to 7~K with a ARS-Helitran crysostat. The room temperature spectrum in the near-infrared to ultraviolet range (4\,000 -- 50\,000\icm) was obtained with a commercial ellipsometer (Woollam VASE). The optical conductivity was obtained by performing a Kramers-Kronig analysis of $R(\omega)$~\cite{Dressel2002}. For the low frequency extrapolation below 40\icm, we used a Hagen-Rubens ($R = 1 - A\sqrt{\omega}$) or superconducting ($R = 1 - A\omega^4$) extrapolation. For the extrapolation on the high frequency side, we assumed a constant reflectivity up to 12.5~eV that is followed by a free-electron ($\omega^{-4}$) response.

%
%
\begin{figure}[tb]
\includegraphics[width=\columnwidth]{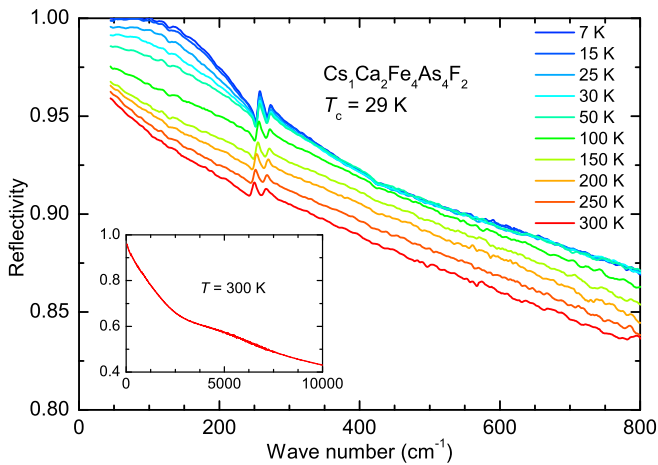}
\caption{ (color online) Far-infrared reflectivity of CsCa$_2$Fe$_4$As$_4$F$_2$ at several temperatures above and below $T_c$. Inset: 300-K reflectivity over a broad frequency range.}
\label{Fig1}
\end{figure}
Figure~\ref{Fig1} displays the in-plane far-infrared reflectivity of CsCa$_2$Fe$_4$As$_4$F$_2$ for several temperatures above and below $T_c$. The inset shows the reflectivity at 300~K for a wide spectral range up to 10\,000\icm. In the normal state, $R(\omega)$ shows a typical metallic response, approaching unity at low frequencies and increasing upon cooling. When entering the superconducting state, the reflectivity below 250\icm\ shows an upturn and reaches a flat, unity response at 7~K, which is a clear signature of the opening of a SC gap. In addition to these gross features, two sharp peaks representing the symmetry allowed in-plane infrared active $E_u$ phonon modes are observed around 250 and 265\icm~\cite{Akrap2009,Xu2015,Yang2017}.

\begin{figure}[tb]
\includegraphics[width=0.98\columnwidth]{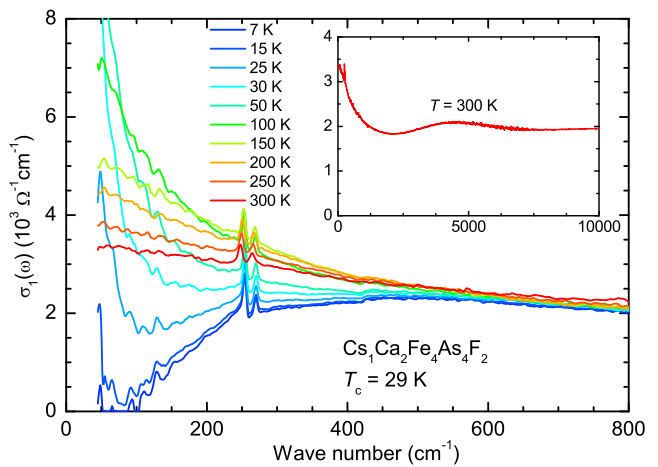}
\caption{ (color online) The real part of the optical conductivity for CsCa$_2$Fe$_4$As$_4$F$_2$ in the far-infrared region at several temperatures above and below $T_c$. Inset: Optical conductivity at 300~K over a broad frequency range.}
\label{Fig2}
\end{figure}
Figure~\ref{Fig2} shows the temperature dependence of the optical conductivity ${\sigma}_1(\omega)$ of CsCa$_2$Fe$_4$As$_4$F$_2$ in the far-infrared above and below $T_c$. The inset shows the 300~K conductivity over a broader spectral range. In the normal state, the optical conductivity has a Drude-like peak centered at zero frequency where the width of the Drude response at half maximum is the value of the quasiparticle scattering rate. As the temperature decreases, this Drude-like peak narrows with a concomitant increase of the low-frequency optical conductivity. Just above $T_c$ at 30~K the Drude peak is quite narrow, suggesting a very small quasiparticle scattering rate at low temperature. Below $T_c$, the formation of the superfluid condensate gives rise to a suppression of $\sigma_1(\omega)$ at low frequencies. This suppression is the consequence of a transfer of spectral weight from finite frequencies to a zero-frequency $\delta(\omega)$ function which represents the infinite dc conductivity of the superconducting condensate that is described by the so-called Ferrel-Glover-Tinkham (FGT) sum rule~\cite{Ferrell1958,Tinkham1959}. For the 7~K spectrum, the low-frequency part of the optical conductivity is strongly suppressed and ${\sigma}_1(\omega)$ vanishes (within the error of the experiment) below $\sim$ 110\icm. Such a complete suppression of the low-frequency optical conductivity is a clear signature of a nodeless SC gap structure as it was also observed in CaKFe$_4$As$_4$~\cite{Yang2017} and in optimally doped Ba$_{1-x}$K$_x$Fe$_2$As$_2$~\cite{Li2008,Dai2013EPL,Xu2017,Mallett2017}.

\begin{figure}[tb]
\includegraphics[width=\columnwidth]{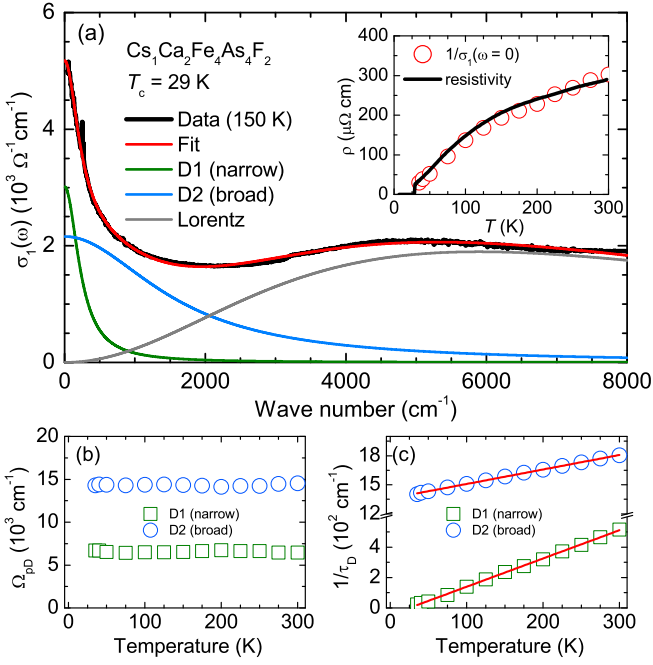}
\caption{ (color online) (a) Optical conductivity of CsCa$_2$Fe$_4$As$_4$F$_2$ up to 8\,000\icm\ (1~eV) at 150~K. The thin red line through the data is the Drude-Lorentz fitting result, which consists of the contributions from a narrow Drude-peak (green line), a broad Drude-peak (blue line), and a Lorentzian (grey line). Inset: Comparison of the dc resistivity, $\rho_{ab}$ (solid line), with the zero-frequency values of the Drude fits to the optical data (open circles). (b) Temperature dependence of the plasma frequency of the Drude-terms $\Omega_{p,D}$. (c) Temperature dependence of the scattering rate $1/\tau_D$ of the Drude terms. The solid red line shows a fit with a $T$-linear behavior.}
\label{Fig3}
\end{figure}
A quantitative analysis of the optical data has been obtained by fitting the ${\sigma}_1(\omega)$ spectra with a   Drude-Lorentz model,
\begin{equation}
\label{DrudeLorentz}
\sigma_{1}(\omega)=\frac{2\pi}{Z_{0}}\biggl[\sum_{j}\frac{\Omega^{2}_{pD,j}}{\omega^{2}\tau_{D,j} + \frac{1}{\tau_{D,j}}} + \sum_{k}\frac{\gamma_{k}\omega^{2}S^{2}_{k}}{(\omega^{2}_{0,k} - \omega^{2})^{2} + \gamma^{2}_{k}\omega^{2}}\biggr],
\end{equation}
where $Z_{0}$ is the vacuum impedance. The first sum of Drude-terms describes the response of the itinerant carriers in the different bands that are crossing the Fermi-level, each characterized by a plasma frequency $\Omega_{pD,j}$ and a scattering rate $1/\tau_{D,j}$. The second term contains a sum of Lorentz oscillators of which each has a resonance frequency $\omega_{0,k}$, a line width $\gamma_k$ and an oscillator strength $S_k$. The corresponding fit to the conductivity at 150~K (thick black line) using the function of Eq.~\ref{DrudeLorentz} (red line) is shown in Fig.~\ref{Fig3}(a) up to 8\,000\icm. As shown by the thin coloured lines, the fitting curve is composed of two Drude terms with small and large scattering rates, respectively, and a Lorentz-term that accounts for the interband transitions at higher energy (grey line). Fits of equal quality have been obtained with this fit configuration for the $\sigma_1(\omega)$ curves at all the measured temperatures in the normal state. The inset of Fig.~\ref{Fig3}(a) compares the temperature dependent values of the dc resistivity $1/\sigma_1(\omega \rightarrow 0)$ deduced from the optical data (open circles) with the ones obtained from the dc transport measurements (solid line). The good agreement confirms the consistency of our modeling of the optical data.

The two-Drude fit indicates that CsCa$_2$Fe$_4$As$_4$F$_2$ has two types of charge carriers with very different scattering rates. A corresponding trend has been reported for various FeSCs for which the narrow (broad) Drude-peak has been assigned to the electron-like (hole-like) bands around the $\Gamma$-point (M-point) of the Brillouin zone~\cite{Wu2010,Dai2013,Nakajima2013}. Fig.~\ref{Fig3}(b) shows the temperature dependence of the plasma frequencies of the two Drude terms which remain constant within the error bar of the measurement, indicating that the band structure hardly changes with temperature. Fig.~\ref{Fig3}(c) displays the temperature dependence of the corresponding scattering rates which both decrease towards low temperature. The red solid line denotes a $T$-linear fit of $1/\tau_D$ that spans the entire temperature range above $T_c$ and is suggestive of a non Fermi-liquid behavior, similar as in optimally doped Ba$_{1-x}$K$_x$Fe$_2$As$_2$~\cite{Dai2013}.

\begin{figure}[tb]
\includegraphics[width=\columnwidth]{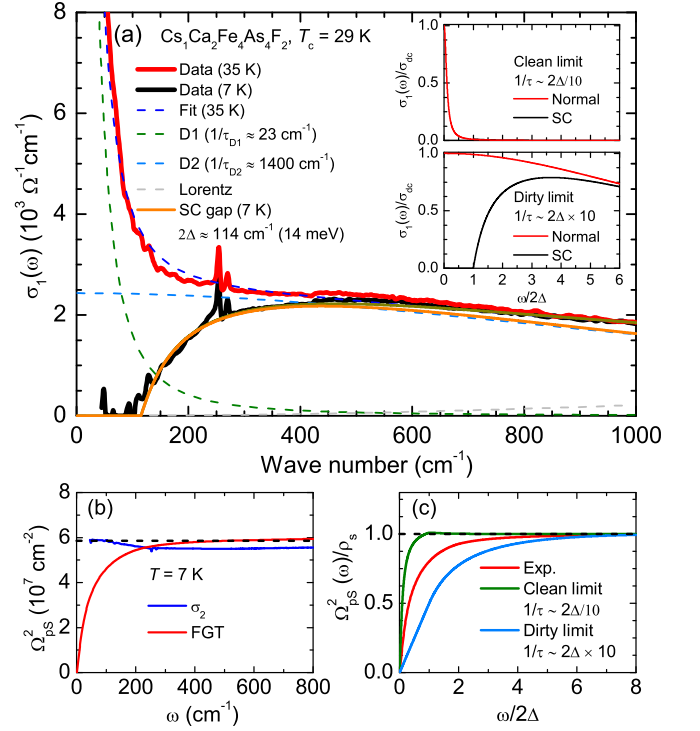}
\caption{ (color online) (a) Optical conductivity of CsCa$_2$Fe$_4$As$_4$F$_2$ above (35~K, thick red line) and below (7~K, thick black line) $T_c$. The dashed lines represent the Drude-Lorentz fits at 35~K. The thin solid line (orange) denotes the dirty-limit Mattis-Bardeen contribution with a superconducting energy gap of $2\Delta \simeq 14$~meV at 7~K. The insets illustrate the superconductivity-induced changes of the optical conductivity in the clean- and dirty-limit cases. (b) The superfluid density $\Omega^2_{pS}$ at 7~K as deduced from the imaginary part of the optical conductivity (blue line) and calculated from the missing spectral weight of the real part according to the FGT sum rule (red line), respectively. (d) Comparison of the measured superfluid density $\Omega^2_{pS}(\omega)$ (red line) and the results calculated for the clean- and dirty-limit cases as discussed in the main text.}
\label{Fig4}
\end{figure}
Next, we focus on the optical response of CsCa$_2$Fe$_4$As$_4$F$_2$ in the superconducting state. Figure~\ref{Fig4}(a) shows the optical conductivity of CsCa$_2$Fe$_4$As$_4$F$_2$ at 35~K just above $T_c$ (thick red line) and at 7~K well below $T_c$ (thick black line). The dashed line shows the fit to the data at 35~K, which consists of a narrow Drude peak with $\Omega_{pD,1} \simeq 6\,500 \pm 200$\icm\ and $1/\tau_{D,1} \simeq 23 \pm5$\icm, and a broad one with $\Omega_{pD,2} \simeq 14\,000 \pm 500$\icm\ and $1/\tau_{D,2} \simeq 1\,400 \pm100$\icm. For the spectrum in the superconducting state at 7~K spectrum, there is a strong suppression of the low-frequency conductivtiy (that sets in below about 400\icm). In particular, as was already noted above, there is a full suppression of the optical conductivity up to a frequency of about 110\icm\ and a steep increase of $\sigma_{1}(\omega)$ toward higher frequency that is the hallmark of a nodeless SC gap with a magnitude of $2\Delta \simeq 14$~meV.

For a quantitative description of the spectrum at 7~K we have used a Mattis-Bardeen-type model of the optical conductivity~\cite{Mattis1958,Zimmermann1991} for each of the two Drude-peaks. Notably, since $1/\tau_{D,1} \ll 2\Delta$ and $1/\tau_{D,2} \gg 2\Delta$, this means that the first (narrow) part of the Drude-response is in the clean limit ($1/\tau_{D} \ll 2\Delta$) whereas the second (broad) part is in the dirty limit ($1/\tau_{D} \gg 2\Delta$). As illustrated in the inset of Fig.~\ref{Fig4}(a), for the former clean-limit case nearly all of the normal state spectral weight is located below $2\Delta$. In the superconducting state, therefore essentially all of this spectral weight is redistributed to the $\delta(\omega)$ function at zero frequency that accounts for the superconducting condensate, leaving no observable conductivity at finite frequency. In the dirty-limit case, much of the normal state spectral weight lies above $2\Delta$ and therefore does not contribute to the SC condensate. The remaining regular part of $\sigma_{1}(\omega)$ has an absorption edge at $2\Delta$ above which the conductivity first rises steeply and eventually levels off and slowly approaches the normal state value above $\omega \sim 6\times2\Delta$. Indeed, as shown by the solid orange line through the data, the $\sigma_1(\omega)$ spectrum at 7~K is rather well described by a single Mattis-Bardeen-type SC gap in the dirty limit. The gap value determined from the fit is $2\Delta \simeq 14$~meV and the ratio $2\Delta/k_BT_c$ is about 4.8, in agreement with a recent heat transport study~\cite{Huang2019}. Therefore, as a consequence of its multiband nature, the optical response in the superconducting state of CsCa$_2$Fe$_4$As$_4$F$_2$ simultaneously satisfies the clean- and dirty-limit conditions.

The magnitude of the zero-frequency $\delta(\omega)$ function and thus the superconducting plasma frequency, $\Omega_{pS}$, and the density of the superfluid condensate, $\rho_s \equiv \Omega^2_{pS}$, can be calculated using the so-called FGT sum rule~\cite{Ferrell1958,Tinkham1959}:
\begin{equation}
\label{eq_FGT}
\Omega^2_{pS} = \frac{Z_0}{\pi^2}\int^{\omega_c}_{0^+}[\sigma_{1}(\omega,T \simeq T_c)-\sigma_{1}(\omega,T \ll T_c)]\, d\omega,
\end{equation}
where $\omega_c$ is chosen such that $\sigma_1(\omega \gtrsim \omega_c)$ is indistinguishable between the normal and superconducting states. Alternatively, the superfluid density is deduced from the imaginary part of the optical conductivity~\cite{Jiang1996} according to:
\begin{equation}
\label{eq_S2}
\rho_{s} = \frac{Z_0}{2\pi}\omega\sigma_{2s}(\omega).
\end{equation}
For equation~\ref{eq_S2} to be accurate, the contribution of the regular (non-superconducting) response to $\sigma_2$ has to be subtracted as described in Refs.~\cite{Dordevic2002,Zimmers2004}.

Figure~\ref{Fig4}(b) shows the superfluid density at 7~K as obtained from the FGT sum rule (red curve) and from $\sigma_2$ (blue curve). The values obtained from both methods agree rather well, as indicated by the black dashed line, and yield a superfluid plasma frequency of $\Omega_{pS} \simeq$ 7\,600\icm\ that corresponds to a magnetic penetration depth of $\lambda = 1/2\pi\Omega_{pS} \simeq210$~nm. The frequency dependence of $\Omega^2_{pS}(\omega)$, normalized to the values of $\rho_s$ and $2\Delta$, is shown by the red line in Fig.~\ref{Fig4}(c). The value of $\Omega^2_{pS}(\omega)$ first increase steeply up to $\omega \sim 2\Delta$ before it starts to saturate and eventually approaches unity around $8\Delta$, indicating that the FGT sum rule is satisfied at this energy scale. For comparison, we also show in Fig.~\ref{Fig4}(c) the corresponding curves obtained assuming that the response of both Drude-bands is in the pure clean-limit (green line) or dirty-limit (blue line). It shows that the pure clean-limit (dirty-limit) case leads to a much steeper (slower) increase of $\Omega^2_{pS}(\omega)$ than in the experimental data, thus confirming our interpretation in terms of a band-selective clean- and dirty-limit scenario.

Overall, the optical response in the superconducting state of the 12442 compound CsCa$_2$Fe$_4$As$_4$F$_2$ compares rather well with the one reported for CaKFe$_4$As$_4$ and optimally doped Ba$_{1-x}$K$_x$Fe$_2$As$_2$, which also show multiple nodeless SC gaps~\cite{Yang2017,Li2008,Dai2013EPL}. The existence of a multiband, clean- and dirty-limit superconductivity has also been reported for FeSCs, such as 11-type FeTe$_{0.55}$Se$_{0.45}$~\cite{Homes2015} and 111-type LiFeAs~\cite{Dai2016}. However, for these latter FeSCs, the description of the SC response required at least two SC gaps that are in the dirty-limit. This may well be the result of a larger disorder, as compared to CsCa$_2$Fe$_4$As$_4$F$_2$. Alternatively, it may indicate that the SC gaps on the three hole-like bands around the $\Gamma$-point, which are responsible for the broad Drude response, have very similar magnitudes in CsCa$_2$Fe$_4$As$_4$F$_2$ but not for the other FeSCs mentioned above.

%
%
To summarize, the optical properties of the multiband superconductor CsCa$_2$Fe$_4$As$_4$F$_2$ ($T_c \simeq 29$~K) have been investigated for numerous temperatures above and below $T_c$. Taking into account the multiband nature of this material, the normal state optical properties have been described by a two-Drude model with a narrow and a very broad Drude peak. In the superconducting state below $T_c \simeq 29$~K, a sharp gap feature is observed in the optical conductivity spectrum with a full suppression of $\sigma_1(\omega)$ below $\sim$ 110\icm\ that is a hallmark of a nodeless superconductor. A quantitative description of the superconducting response has been obtained with a two-gap model that assumes the coexistence of a clean-limit gap on the narrow Drude band and a single dirty-limit gap on the broad Drude band. For the latter we derived a magnitude of $2\Delta \simeq 14$~meV. The overall plasma frequency or condensate density has been deduced as $\Omega_{pS} \simeq$ 7\,600\icm, corresponding to a magnetic penetration depth of 210~nm.

%
%
Work at the University of Fribourg was supported by the Schweizer Nationalfonds (SNF) by Grant No. 200020-172611.

%

\end{document}